\documentclass[aps,twocolumn,groupedaddress,showpacs]{revtex4}
\usepackage{graphicx}

\begin{document}

\title{Excitations in the halo nucleus $^{6}$He following the
$^7$Li($\gamma,p$)$^6$He reaction}

\author{M. J. Boland}
\author{M. A. Garbutt}
\author{R. P. Rassool}
\author{M. N. Thompson}
\author{A. J. Bennett}
%\thanks{}
\affiliation{School of Physics, The University of Melbourne, Victoria
3010, Australia}

\author{J. W. Jury}
\affiliation{Trent University, Peterborough, Ontario, Canada K9J 7B8}

\author{J.-O. Adler}
\author{B. Schr\"oder}
\author{D. Nilsson}
\author{K. Hansen}
\author{M. Karlsson}
\author{M. Lundin}
\affiliation{Department of Physics, University of Lund, P.O. Box 118, S-221 00 Lund, Sweden}

\author{I. J. D. MacGregor}
\affiliation{Department of Physics and Astronomy, University of
Glasgow, Glasgow G12 8QQ, Scotland}

% body of paper here

\date{\today}

\begin{abstract}
A broad excited state was observed in $^6$He with energy $E_{x} = 5
\pm 1$ MeV and width $\Gamma = 3 \pm 1$ MeV, following the reaction
$^7$Li($\gamma,p$)$^6$He.  The state is consistent with a number of
broad resonances predicted by recent cluster model calculations.  The
well-established reaction mechanism, combined with a simple and
transparent analysis procedure confers considerable validity to this
observation.
\end{abstract}

% see http://www.aip.org/pacs/pacs99/

\maketitle

% body of paper here
The physics of nuclei approaching the neutron drip-line is of interest
as a means of further refining our understanding of the
nucleon-nucleon potential. Amongst these so-called ``halo'' nuclei,
$^{6}$He has received considerable attention.  The established level
structure of $^{6}$He \cite{Ajzenberg:88} has been questioned for some
years in a number of theoretical calculations.  These considered
extended neutron distributions by modeling $^6$He as a $^4$He+n+n
three-body cluster.  A common feature of these calculations is
low-lying structure, above the well known $2^{+}$ first excited state.
The nature of this structure was initially thought to be a soft dipole
resonance \cite{Hansen:87,Suzuki:91}, with two halo neutrons
oscillating against the core. However, more recent calculations refute
this and postulate that it is caused by three-body dynamics
\cite{Csoto:94,Danilin:98,Thompson:00}.

Experimental measurements on the $^{6}$He system have so far been
concentrated on charge exchange reactions of the type
$^{6}$Li($^{7}$Li,$^{7}$Be)$^{6}$He
\cite{Sakuta:93,Janecke:96,Annakkage:99,Nakayama:00} and
$^{6}$Li($t$,$^{3}$He)$^{6}$He \cite{Nakamura:00}. All these results
have reported low-lying strength in the reaction cross section at
roughly the predicted energies by calculations, but none are able to
determine the nature of the observed structure.

In each case the analysis of these experiments has involved several
controversial assumptions in the background removal process.  In
particular, the non-resonant background in the ($^{7}$Li,$^{7}$Be)
reaction was calculated but not measured.  This process must include
degrees of freedom due to the excited states of both the projectile
and the ejectile.  In one case \cite{Nakayama:00}, non-resonant
background contributions to the cross section were not included at
all.

Background subtraction is only one of the complications involved with
heavy-ion reactions.  Another difficulty is that many possible
combinations of angular-momentum transfer exist between projectile and
target.  One of the simplest charge exchange reactions, namely
$(n,p)$, does not suffer the same problem. However, the poor
resolution of these $(n,p)$ experiments makes it difficult to see even
the commonly resolved $2^{+}$ state.  Reactions of the type
$(t,^{3}$He) also suffer from poor resolution, and use the same
background removal process as the ($^{7}$Li,$^{7}$Be) reactions
\cite{Nakamura:00}.  In contrast, tagged photon measurements have a
relatively simple and unambiguous background removal procedure that is
proven and well established
\cite{Owens:90,MacGregor:91,VanHoorebeke:92b,Kuzin:98} (and references
therein).

This paper reports the presence of a broad resonance at an excitation
energy of 5 MeV in $^{6}$He that has been observed following the
$^{7}$Li($\gamma,p$)$^{6}$He photonuclear reaction. The measurement
was made in the energy range of $E_{\gamma} =$ 50--70 MeV, using the
MAX-lab tagged photon facility \cite{Adler:97} at Lund University.
The protons and other charged particles were detected with solid-state
spectrometers, each consisting of a thick HP-Ge $E$-detector and a
thin Si $\Delta E$-detector.  These were placed at angles of $\theta =
30^{\circ}, 60^{\circ}, 90^{\circ}, 120^{\circ}$, and $150^{\circ}$ to
the photon beam, similar to the configuration described in
\cite{Dias:95}.  A 1 mm thick target of 99.9\% pure $^{7}$Li was
placed at $60^{\circ}$ to the photon beam.

\begin{figure}[ht]
 \includegraphics{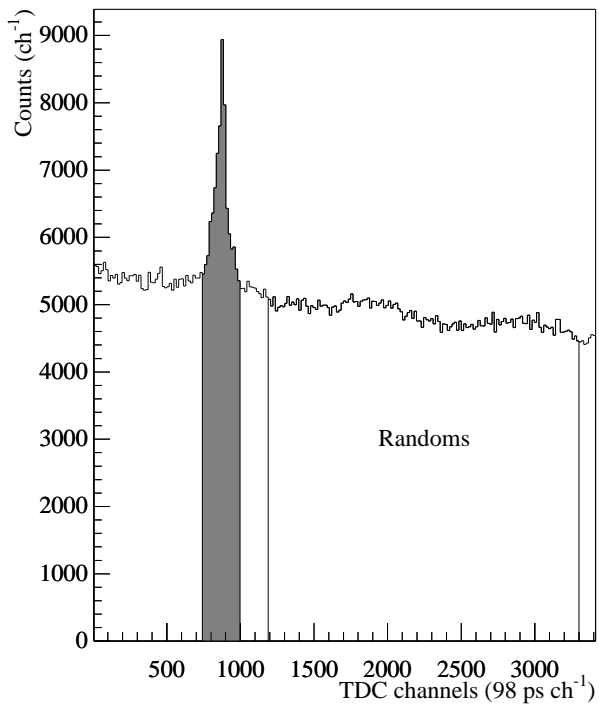}
 \caption{The time correlation spectrum between protons and tagged
 photons for $\theta = 60^{\circ}$. The 6 ns wide prompt peak (shaded)
 is clearly visible on top of a random background (labeled).}
 \label{fig:tdc}
\end{figure}

\begin{figure}[ht]
 \includegraphics{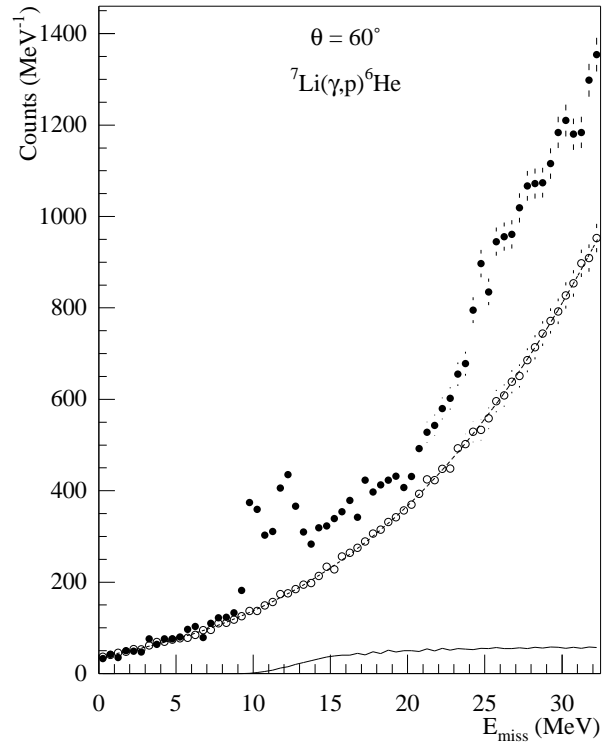}
 \caption{Proton missing-energy spectrum at $\theta = 60^{\circ}$
 showing (i) the random background (open dots) with a polynomial fit
 (dotted line) (ii) the calculated $(\gamma,pn)$ background (solid
 line) and (iii) the prompt protons (filled dots).} 
 \label{fig:emissbg}
\end{figure}

\begin{figure}[ht]
 \includegraphics{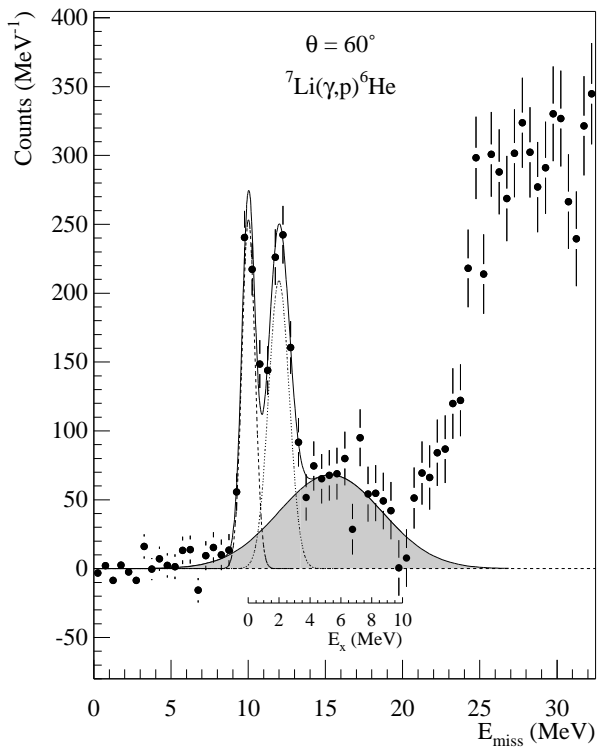}
 \caption{Proton missing-energy spectrum at $\theta = 60^{\circ}$
 following the reaction $^{7}$Li($\gamma,p$)$^{6}$He with the
 background contributions subtracted. The $^{6}$He excitation energy
 scale is drawn on for reference.} 
 \label{fig:emiss}
\end{figure}

Protons were selected from other charged particle events by use of a
particle-identification plot of the energy lost in the full-energy
detector, versus that lost in the $\Delta E$-detector.  Protons
correlated with tagged photons were located in a narrow \emph{prompt}
timing peak, shown shaded in Fig.~\ref{fig:tdc}, sitting on a timing
spectrum of random events.  Missing-energy spectra were produced from
a cut on the prompt peak at each angle (filled dots in
Fig.~\ref{fig:emissbg}). The missing energy is defined as $E_{miss} =
E_{\gamma} - T_{p} - T_{R}$, where $T_{R}$ is the kinetic energy of
the $^{6}$He nucleus, and $T_{p}$ is the kinetic energy of the emitted
proton. The excitation energy, shown in Fig.~\ref{fig:emiss} is
related to $E_{miss}$ by $E_{x} = E_{miss} - Q$, where $Q$ is the
proton separation energy, and for the reaction
$^{7}$Li($\gamma,p$)$^{6}$He, $Q$ = 10.0 MeV.  The contribution of
random proton events in the prompt region, was measured by making a
cut on the random background region (labeled in Fig.~\ref{fig:tdc}).
The resulting featureless background spectrum (open circles in
Fig.~\ref{fig:emissbg}) was normalised and fitted, before being
subtracted from the spectrum of the prompt region.

The contribution due to the $(\gamma,pn)$ reaction (threshold
$E_{miss}$ = 11.9 MeV) also needed to be considered.  The momentum
distribution of this background channel was calculated using a
Monte-Carlo model of direct two-nucleon emission \cite{McGeorge:95},
that included all the experimental parameters, and covered the full
phase-space of the experiment.  The peak of the $(\gamma,pn)$
missing-energy distribution is located at $E_{miss} = 29$ MeV (see
Fig.~\ref{fig:emissbg}) and as such cannot account for all the
strength observed between $E_{miss} =$ 3--10 MeV. The $pn$-background
was normalised in a consistent manner for all angles, then subtracted
such that the net missing-energy spectrum was positive at all
energies.  The resulting missing-energy spectrum of protons emitted at
$\theta = 60^{\circ}$ is shown in Fig.~\ref{fig:emiss}.

Protons leading to the ground state and the first excited state at
$E_{x} =$ 1.8 MeV can be clearly seen. Evidence for the known second
excited state near $E_{x} \sim$ 14 MeV can be distinguished at the
onset of the high missing-energy region of the spectrum.
Significantly, the evidence for a broad state can be seen in the
region between $E_{x} \sim$ 3--10 MeV.  A fit of three Gaussians to
the data in Fig.~\ref{fig:emiss} gives a width of $\Gamma = 3 \pm 1$
MeV and a centroid energy of $E_{x} = 5 \pm 1$ MeV to the new
structure, on the assumption that it is a single resonance.

The present experiment, like those using charge exchange reactions, is
unable to define the exact nature of the observed resonance.  The
strongest candidates seem to be a $1^{-}$ soft dipole mode and a
second 2$^{+}$ state, predicted by Suzuki \cite{Suzuki:91} and others
\cite{Zhukov:92,Danilin:97,Ershov:97,Danilin:98,Ershov:00}. A
calculation of the the $E1$ breakup of $^6$He \cite{Thompson:00} shows
an enhancement to the $1^-$ continuum at an energy consistent with the
measurement presented here.  It is possible that the strength we
observe in the $^7$Li($\gamma,p$)$^6$He cross section at 5 MeV is
evidence of the $1^-$ dipole and the positive parity states, both of
which were predicted by Danilin {\it et al} \cite{Danilin:98}. A
complete analysis of our data, including the angular distribution, may
clarify the nature of the structure and thereby validate some of the
model assumptions.

%%\bibliographystyle{apsrev}
%%\bibliography{/home/markjb/thesis/photonuc}

\end{document}